\documentclass[graybox]{svmult}

\usepackage{type1cm}        
\usepackage{makeidx}         
\usepackage{graphicx}        
\usepackage{multicol}        
\usepackage[bottom]{footmisc}

\usepackage{newtxtext}       %
\usepackage[varvw]{newtxmath}       

\newcommand*{\La}{\cal{L}}
\newcommand*{\no}{\noindent}
\newcommand*{\bea}{\begin{eqnarray}}
\newcommand*{\eea}{\end{eqnarray}}
\newcommand*{\be}{\begin{equation}}
\newcommand*{\ee}{\end{equation}}
\newcommand*{\pd}{\partial}
\newcommand*{\pdm}{\pd_{\mu}}
\newcommand*{\pdn}{\pd_{\nu}}

\newcommand*{\pref}[1]{(\ref{#1})}

\newcommand*{\mn}{{\mu\nu}}
\newcommand*{\rr}{\mathbb{R}}

\newcommand*{\prefr}[2]{(\ref{#1}-\ref{#2})} 
\newcommand*{\nn}{\nonumber}

\newcommand*{\tr}{\mathrm{tr}}

\newcommand{\bma}{\begin{pmatrix}}
\newcommand{\ema}{\end{pmatrix}}

\newcommand*{\op}{{{\cal O}}}
\newcommand*{\la}{\left\langle}
\newcommand*{\ra}{\right\rangle}

\makeindex             

\begin{document}

\title*{The Fr\"ohlich-Morchio-Strocchi mechanism: A underestimated legacy}
\author{Axel Maas}
\institute{Axel Maas \at Institute of Physics, NAWI Graz, University of Graz, Universit\"atsplatz 5, 8010 Graz, \email{axel.maas@uni-graz.at}}
%
%
\maketitle

\abstract{There is an odd tension in electroweak physics. Perturbation theory is extremely successful. At the same time, fundamental field theory gives manifold reasons why this should not be the case. This tension is resolved by the Fr\"ohlich-Morchio-Strocchi mechanism. However, the legacy of this work goes far beyond the resolution of this tension, and may usher in a fundamentally and ontologically different perspective on elementary particles, and even quantum gravity.}

\section{Introduction}

Non-Abelian gauge theories of Yang-Mills type \cite{Yang:1954ek,Bohm:2001yx,Montvay:1994cy}, no matter the matter content, have a highly interesting feature. They are based on gauge (Lie-)groups, which do not form simple manifolds. This has far reaching consequences. Probably the most important one is that it is not possible to introduce global coordinate systems \cite{Singer:1978dk,Gribov:1977wm,Lavelle:1995ty}, an issue known as the Gribov-Singer ambiguity. This feature stems from the group structure, and is thus independent of the parameters of the theory, especially the value of any coupling constants. On top of this, non-Abelian gauge theories are affected, as any other quantum field theory, by Haag's theorem \cite{Haag:1992hx}, which implies that a non-interacting theory and an interacting one are not unitarily equivalent. In a Yang-Mills theory, this is amplified by the fact that the free gauge theory would have a different character, namely to be reduced to $N_g$ non-interacting Abelian gauge theories, where $N_g$ is the number of generators of the gauge group.

This appears to imply  that a conventional perturbative treatment \cite{Bohm:2001yx} should not be possible at all. The elementary particles cannot act as asymptotic states due to Haag's theorem. And the required gauge-fixing for the employed saddle-point approximation in perturbation theory is not well-defined due to the Gribov-Singer ambiguity.

This appears to have "just" the consequence that genuine non-perturbative methods are required, and especially non-trivial asymptotic states are needed. The prime example is QCD. Here, asymptotic states are hadrons, and perturbation theory can at best be applied in special kinematics, where the involved field amplitudes are small enough that the group manifold is only probed within a single patch. Due to the strong coupling, however, this is generally agreed upon anyways \cite{Bohm:2001yx,Montvay:1994cy,Dissertori:2003pj,BeiglboCk:2006lfa}.

In the weak interactions, the situation is, on conceptual grounds, the same \cite{Montvay:1994cy,Osterwalder:1977pc,Fradkin:1978dv,Banks:1979fi,Frohlich:1980gj,Frohlich:1981yi}. Thus, it is not surprising that, e.\ g., there is no qualitative distinction between the strong-coupling case and the weak coupling case, as they are analytically connected states of the theory \cite{Osterwalder:1977pc,Fradkin:1978dv}. But ignoring these fundamental questions and just applying perturbation theory turns out to be extremely successful in describing experimental results to high quantitative precision \cite{Bohm:2001yx,pdg}. This is attributed to the Brout-Englert-Higgs (BEH) effect \cite{Englert:1964et,Higgs:1964pj,Higgs:1964ia,Higgs:1966ev,Guralnik:1964eu,Kibble:1967sv,Englert:1966uea}: It 'breaks' the gauge symmetry, effectively turning it into a non-gauge theory which does not need to take care of these issue. But, formally, a gauge symmetry cannot be broken by virtue of Elitzur's theorem \cite{Elitzur:1975im}, nor does this alleviates Haag's theorem.

It is precisely here, where Giovanni Morchio's legacy in form of the Fr\"ohlich-Morchio-Strocchi (FMS) mechanism \cite{Frohlich:1980gj,Frohlich:1981yi} is the decisive puzzle piece. It explains how both aspects, the phenomenological success and the formal insights, can both be correct at the same time. How this happens will be discussed in section \ref{s:fms}. But while the original papers \cite{Frohlich:1980gj,Frohlich:1981yi} were mainly concerned with resolving this paradox, the legacy and implications of this work transcends in its importance the resolution of the paradox by far. In fact, it creates a framework, the FMS framework, to deal with a quite large class of theories effectively.

In the following, the FMS framework and the FMS mechanism, and some of their consequences, are presented, as there are:
\begin{itemize}
 \item Experimental testability of the field theoretical underpinnings, section \ref{s:pheno}.
 \item Consequences for non-Abelian Yang-Mills-Higgs theories beyond the standard model, section \ref{s:bsm}.
 \item Applications beyond Yang-Mills-Higgs theories, section \ref{s:bym}.
 \item Ontological implications, section \ref{s:phil}.
\end{itemize}
In fact, the FMS mechanism, and the formal aspects on which it is build, have the potential to fundamentally transform our view of 'elementary' particle physics \cite{Berghofer:2021ufy}, and thus the way how we perceive reality. While the need to take gauge invariance seriously has been pointed out repeatedly before, and in fact on formal grounds \cite{Singer:1978dk,Gribov:1977wm,Haag:1992hx,Osterwalder:1977pc,Fradkin:1978dv,Banks:1979fi,Elitzur:1975im,'tHooft:1979bj}, it has been the work of Morchio and his collaborators \cite{Frohlich:1980gj,Frohlich:1981yi} to show how the subtleties work out in practice. They thereby paved the way for a more holistic picture of gauge symmetries, and how they are (not) relevant \cite{Berghofer:2021ufy}.

Given all these implications, it appears surprising that this has found so far no entry even in specialized textbook, much less has become the standard approach. Especially as the necessary additional effort is at best moderate, see section \ref{s:fms}. And the original papers \cite{Frohlich:1980gj,Frohlich:1981yi} are now more than 40 years old. While a full historical and sociological investigations is not (yet) available, superficial investigations \cite{Maas:2017wzi} show that the insidious combination of the properties of the standard model and the success of the FMS mechanism itself appear to be the reason for that. Because in the particular case of the standard model, the FMS mechanism explains why only slight deviations can be expected, compared to a perturbative treatment. In fact, so slight, that they have not yet been observable in experiment, see section \ref{s:pheno}. As a consequence, its additional layer of complexity has not been needed, as perturbation theory alone was sufficient. Thus, it got almost forgotten, and the (formally incorrect) idea of gauge symmetry breaking by the BEH effect has become accepted lore. Only within the philosophy of science community the challenged posed to our understanding by Elitzur's theorem and its contradiction to the BEH effect has remained a matter of importance \cite{Berghofer:2021ufy,Friederich:2013,Friederich:2014,Lyre:2004,Lyre:2008af,Francois:2018,Smeenk:2006,Struyve:2011nz}. Especially, within the philosophy of gauge symmetry literature, even disbelief about the treatment of the issue by physicists was expressed.

Turning the whole story around, there is an important discovery awaiting. Either we are able to experimentally discover the correctness of the consequences of the FMS mechanism, or not. In the former case, this will make the FMS mechanism the accepted approach for treating the BEH effect, and will have far-reaching consequences for model building \cite{Maas:2017wzi}, see sections \ref{s:bsm} and \ref{s:bym}. Or, this will disprove our fundamental understanding of quantum gauge theories, as encoded in \cite{Singer:1978dk,Haag:1992hx,Osterwalder:1977pc,Fradkin:1978dv,Frohlich:1980gj,Frohlich:1981yi,Elitzur:1975im,Maas:2017wzi}, sending us back to the drawing board, and perhaps open entirely new avenues. As the effects are predictable and entirely fixed by the known parameters of the standard model, this decision can be performed. Even if it is a formidable, though manageable, task, see section \ref{s:pheno}.

The only thing, which is not an option, is to ignore this tension. Because if the understanding of quantum gauge theories is correct, the consequences of ignoring the tension could easily be mistaken for signatures of physics beyond the standard model \cite{Maas:2017wzi,Maas:2013aia,Maas:2015gma,Jenny:2022atm}. Indeed, there is an off-chance that this may have already happened \cite{Egger:2017tkd}.

\section{The FMS mechanism}\label{s:fms}

The starting point of the FMS framework is a simple statement. Given any expression $\op$, which transforms in a linear representation of a (continous) non-Abelian\footnote{In fact, similar arguments do hold also in the Abelian case \cite{Lenz:1994tb,Dudal:2021pvw,Dudal:2019pyg}, and are then augmented by the usual subtleties of Abelian gauge theories \cite{Haag:1992hx}. This will not be detailed here, but follows along very similar lines, including the confirmation in lattice simulations \cite{Woloshyn:2017rhe,Lewis:2018srt}.} gauge group $G$, an invariant group measure ${\cal D}\mu$, and an invariant action as weight factor $\exp(iS)$, it follows\footnote{The original work \cite{Frohlich:1980gj,Frohlich:1981yi} used a lattice regularization in Euclidean space-time to carefully bound expressions. While important on a formal level, this turns out to be transparent to the following, and therefore will be suppressed. Especially, behavior at sufficiently low energies compared to a (lattice) cutoff is likely independent on such details \cite{Seiler:1982pw,Hasenfratz:1986za}.} that \cite{Frohlich:1980gj,Frohlich:1981yi}
\be
\int{\cal D}\mu \op e^{iS}=0\label{sym},
\ee
\no because 
\be
\la \op\ra=\int{\cal D}\mu \op e^{iS}=\int{\cal D}\mu^{g^{-1}}\op e^{iS}=\int{\cal D}\mu \op^g e^{iS}=\la \op^g\ra\nn,
\ee
\no where $g$ is a gauge transformation. This can only be true for arbitrary $g$ if $\la\op^g\ra=\la \op\ra=0$. This statement is a generalization of Elitzur's theorem \cite{Elitzur:1975im}. Thus, there can be no spontaneous breaking of a gauge symmetry by formation of a gauge-dependent condensate like in the BEH effect, which in turn would break the gauge symmetry. Thus, the gauge symmetry remains unbroken.

Additionally, this approach closes a loop hole in the original derivation, which assumed analyticity of the free energy in external sources, which is not necessarily the case \cite{Maas:2013sca}. In fact, this statement also applies to global groups \cite{Maas:2017wzi,Sartori:1992ib}. As a consequence, expectation values of gauge-dependent quantities necessarily vanish, if the gauge symmetry is unbroken. But this can then happen only by gauge-fixing \cite{Frohlich:1980gj,Frohlich:1981yi,Elitzur:1975im}.

Thus, without breaking gauge symmetry explicitly by gauge fixing, it remains necessarily unbroken. The BEH effect is therefore not a physical effect, but rather only a particular useful gauge choice implemented by, e.\ g., the 't Hooft-$R_\xi$ gauges \cite{Lee:1974zg}. As a consequence, the Higgs vacuum expectation value is introduced by the gauge-fixing and thus gauge-dependent. Its actual value needs still to be determined from the gauge-fixed quantum effective potential, and whether it can be non-zero remains a dynamical, albeit gauge-dependent \cite{Caudy:2007sf,Maas:2012ct,Dobson:2022ngz,Kajantie:1998yc}, question.

As a consequence, the Gribov-Singer ambiguity still applies, and thus the classification of physical states using BRST symmetry fails \cite{Fujikawa:1982ss}. Rather, fully and manifestly gauge-invariant operators are needed to construct asymptotic states \cite{Banks:1979fi,Frohlich:1980gj,Frohlich:1981yi}. Fortuitously, this also elegantly satisfies Haag's theorem, as the asymptotic states are no longer necessarily non-interacting elementary particles.

While this is a field-theoretically satisfying prescription, this implies effectively to work with bound states. While non-perturbative methods exist to do so, they are much more demanding than perturbation theory. They work very well for theories like QCD \cite{Montvay:1994cy,BeiglboCk:2006lfa,Brambilla:2014jmp,Gross:2022hyw}. But the large hierarchy of the standard model, covering at least twelve orders of magnitude, make them practically not (yet) applicable. Moreover, the CP-breaking character of the weak interactions poses still conceptual challenges for some of them \cite{Maas:2017wzi,Hasenfratz:2007dp}.

It is here, where the second part of the FMS framework becomes central, the FMS mechanism \cite{Frohlich:1980gj,Frohlich:1981yi}. It can be argued that the Gribov-Singer ambiguity is quantitatively not important in the presence of a BEH effect \cite{Lenz:2000zt}, for which at least some circumstantial evidence exists \cite{Maas:2010nc,Capri:2013oja}. Similarly, the success of perturbation theory \cite{Bohm:2001yx,pdg} ignoring all of these issues requires understanding, but indicate that there exists some suppression mechanism.

The FMS mechanism now utilizes that any kind of perturbation theory is indeed also a small field-amplitude expansion \cite{Rivers:1987hi}. Thus, if the dominating field configurations in the path integral are characterized by small-field amplitude fluctuations around some fixed field configuration, an expansion should be still quantitatively good. This could happen, e.\ g., due to a BEH effect, where the Higgs field develops after gauge fixing a vacuum expectation value as dominating field configuration. Hence, it should be possible to still expand accordingly, i.\ e.\ performing a saddle-point expansion around the Higgs vacuum expectation value. However, following the FMS framework, the expansion needs to be performed around the correct asymptotic states, which are manifestly gauge-invariant.

Consider as an example the simplest theory having all of these features, the Higgs sector of the standard model. Its Lagrangian is \cite{Bohm:2001yx}
\bea
\La&=&-\frac{1}{4}\tr W_\mn W^\mn+(D_\mu X)^\dagger D^\mu X+V(\det X)\label{higgss}\\
W_\mn&=&\pd_\mu W_\nu-\pd_\nu W_\mu+ig\left[W_\mu,W_\nu\right]\nn\\
D_\mu&=&\pd_\mu+g W_\mu\nn
\eea
\no where $g$ is the gauge coupling, $W_\mu=\tau^a W_\mu^a$ are algebra-valued gauge fields with the generators of the gauge group $\tau^a$, in the standard model SU(2). $X$ is the matrix-valued Higgs field derived from the complex doublet $h$ \cite{Maas:2017wzi}. This form makes explicit that the Higgs field is in the fundamental representation of the gauge group under left-multiplications, and also in a fundamental representation with respect to a right-multiplication of an additional global SU(2) group. It should be noted that $X$ itself is not SU(2)-valued. The potential $V$ is required to be invariant under both symmetries, which is ensured by construction.

The BEH effect is made possible by a suitable gauge-fixing, which explicitly breaks the gauge symmetry completely \cite{Bohm:2001yx,Maas:2017wzi}. After gauge-fixing, the Higgs field is then split conveniently as
\be
X(x)=V+\eta(x)\label{behsplit}
\ee
\no where $V$ is a constant. It is convenient, but not necessary, to choose $V=v\mathbf{1}$. If the quantum-effective action allows for $v\neq 0$, a BEH effect takes place\footnote{It is important to note that this is not \cite{Frohlich:1980gj,Frohlich:1981yi,Maas:2017wzi} a background-field approach \cite{Bohm:2001yx}, as the splitting happens after gauge fixing and not before. Especially, there is no additional classical gauge symmetry of the split-off field. The previous gauge-fixing has already broken the gauge symmetry.}.

The FMS framework demands to formulate matrix elements of physical observables in terms of gauge-invariant operators. To interpret them as particles in terms of asymptotic states requires them to be local. Local, manifestly gauge-invariant operators in a non-Abelian gauge theory are necessarily composite \cite{Lavelle:1995ty,Maas:2017wzi}. Especially, this implies the split \pref{behsplit} is not applied at the level of the Lagrangian, like in perturbation theory. Rather, the FMS mechanism works by first writing down the desired matrix element in terms of local, composite operators and only then the split \pref{behsplit} is applied. Of course, the local composite operators can still carry global quantum numbers, especially spin or the those from the global SU(2) symmetry in \pref{higgss}.

The simplest case is the propagator of a scalar singlet. A suitable operator\footnote{At first sight, this appears to be the same as the fields appearing in unitary gauge \cite{Bohm:2001yx}. However, unitary gauge introduces a non-trivial Faddeev-Popov operator due to gauge defects, and is thus different \cite{Maas:2017wzi}.} would be $\det X$. The simplest, non-trivial, matrix element of this operator is the propagator. Taking only the connected part yields
\bea
\la \det X(x)\det X(y)\ra&=&v^2\la\tr\eta(x)\tr\eta(y)\ra\label{fms1}\\
&&+v\la\det \eta(x)\tr\eta(y)+\tr\eta(x)\det \eta(y)\ra+\la\det\eta(x)\det\eta(y)\ra.\nn\\\label{fms2}
\eea
\no Since the left-hand side is gauge-invariant, so the sum on the right-hand side needs to be. However, the individual terms are not necessarily so. So far, this is an exact rewriting, and thus not a priori progress.

Because any kind of perturbative expansion is assuming that the quantities are analytic in the expansion parameter, a perturbative expansion cannot spoil the gauge-invariance of the right-hand side, if order-by-order the sum in powers of $v$ is kept \cite{Maas:2020kda}. The first term \pref{fms1} is the propagator of the elementary Higgs particle. Dropping the remaining terms \pref{fms2} and expanding the elementary Higgs propagator in a perturbative series in the couplings yields that the bound-state propagator in this approximation is identical to the elementary Higgs propagator to all orders in perturbation theory. Especially, the poles coincide and thus the bound state mass is the same as the elementary one to all orders in perturbation theory\footnote{At loop order, it is necessary to chose a pole scheme to fix the pole position, which is independent of the gauge parameter by virtue of the Nielsen identities \cite{Maas:2017wzi,Nielsen:1975fs}.}. Of course, at loop-level the elementary Higgs propagator is gauge-dependent \cite{Maas:2020kda,Dudal:2020uwb}. This was expected, as only the sum would be gauge-invariant. This gauge-invariance of the full sum has indeed been demonstrated explicitly at one-loop order \cite{Maas:2020kda,Dudal:2020uwb}. Furthermore, it was seen explicitly that only at energy scales much larger than $v$ deviations from ordinary perturbative result arise. This explicitly shows how the perturbative success can be recovered by the FMS mechanism, at least in principle: The term \pref{fms1} dominates over \pref{fms2} in the experimentally probed regime. This is how the FMS mechanism resolves the paradox: The composite state has a FMS-dominant contribution, which coincides with one of the elementary particles.

Of course, this is only a self-consistency statement. While the agreement with experiment is certainly strong support, explicitly evaluation of both sides with non-perturbative methods provides another stringent tests. This has been done using lattice methods, confirming the FMS mechanism, see \cite{Maas:2017wzi} for a review. Especially, no additional poles due to further bound states or resonances are observed below the inelastic threshold in this theory in all investigated channels, confirming the experimental findings.

Another important structure is visible when considering the triplet vector channel. Consider the following operator and its FMS expansion
\be
\tr\tau^i X^\dagger D_\mu X=v^2\delta^i_a W_\mu^a+...\label{wfms}
\ee
\no Thus, this gauge-invariant vector operator, which is a triplet under the global SU(2) symmetry carried by the Higgs field, is mapped to the gauge boson field. With the same reasoning as before, this yields that the mass of the composite physical particle is the same as the one of the elementary particle. The $W$ is hence the FMS-dominant contribution of the composite state. More importantly, there is a matrix $c^i_a=\delta^i_a$, carrying both global group $i$ and local gauge group $a$ indices, which yields a map of the global triplet to the gauge triplet. This is the mechanism by which the degeneracy from the gauge fields is transported to the gauge-invariant physical states. This mapping will be of special importance in section \ref{s:bsm}. This structure was again confirmed at one-loop order \cite{Dudal:2020uwb} and on the lattice \cite{Maas:2017wzi}.

\begin{figure}
\begin{minipage}[c]{0.8\textwidth}
 \includegraphics[width=\textwidth]{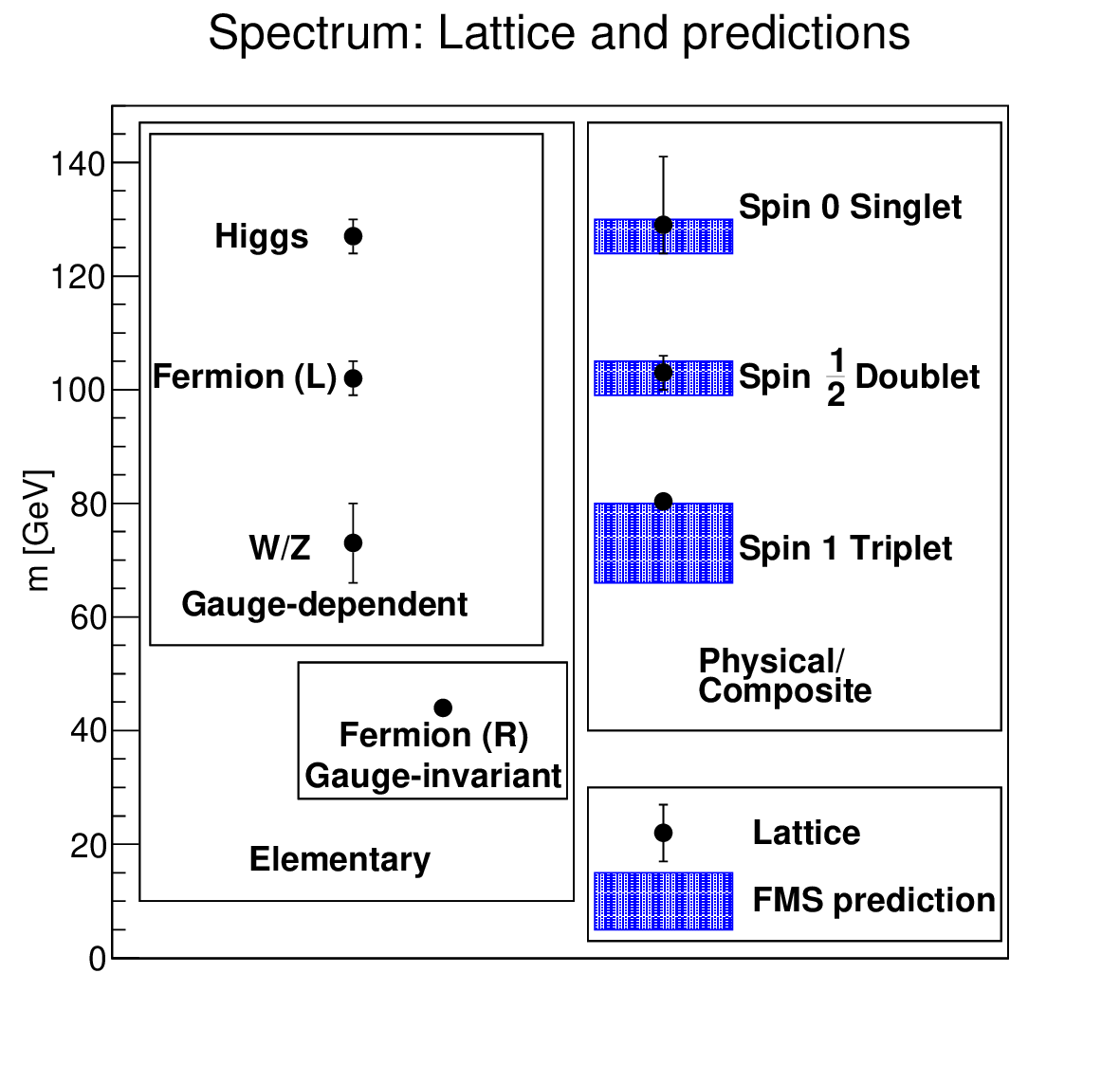}
 \end{minipage}\hfill
 \begin{minipage}[c]{0.2\textwidth}
 \caption{A sample lattice spectrum for quenched vectorial fermions compared to the predictions from the FMS mechanism, alongside the elementary particle properties \cite{Afferrante:2020fhd}. Note that a right-handed ungauged fermion is in addition necessary to construct in such a theory power-counting renormalizable Yukawa interaction with the Higgs \cite{Afferrante:2020fhd}.}\label{spec}
\end{minipage}
\end{figure}

The whole construction can be repeated for the remainder of the standard model. This includes left-handed leptons \cite{Frohlich:1980gj,Frohlich:1981yi,Maas:2017wzi,Egger:2017tkd,Shrock:1986bu} and hadrons \cite{Maas:2017wzi,Egger:2017tkd}. It turns out that only the very special structure of the standard model ensures the correct assignment of degeneracies and further quantum numbers like electric charge. Furthermore, always the physical, gauge-invariant states map to elementary states, if the latter exist, and to scattering states otherwise. Unfortunately, as noted before, the whole standard model cannot be yet treated non-perturbatively. Within simplified models of the lepton sector, however, also agreement is found \cite{Afferrante:2020fhd}, see figure \ref{spec}. Hence, all possible theoretical tests of the FMS mechanism so far have confirmed it. Not to mention that it explains why perturbation theory and experiment agree so well. It should be noted in passing that the FMS mechanism implies at tree-level or in a pole scheme that most bound states have mass defects between 50\% (for the Higgs) to 75\% (for weak gauge bosons) to almost 100\% (electrons and left-handed neutrinos). It is thus a highly relativistic effect, not accessible  \cite{Maas:2017wzi,Maas:2012tj} to quantum-mechanical models or heavy-particle effective field theories such as e.\ g.\ in \cite{Grifols:1991gw}.

All of this resolves the contradiction between the field-theoretical arguments and the success of perturbation theory, at least on the theoretical level.

\section{Phenomenological implications and experimental tests}\label{s:pheno}

Of course, it is not sufficient that analytical expressions and lattice results coincide. To ensure that this is actually the correct description of the standard model, experimental tests are needed. This needs to detect experimentally consequences of additional terms like \pref{fms2}. Since experiments usually involve scattering, it is required to address the scattering of composite states.

This is not a conceptual issue per se, as composite state scattering can be addressed within the LSZ formalism in the same way as the scattering of elementary states \cite{Bohm:2001yx,Haag:1992hx,Meissner:2022odx}. In particular, the choice of asymptotic operators does not matter.

But it furthermore needs a possibility to calculate such processes, as the appearing matrix elements and wave functions are intricate objects. Fortunately, the FMS mechanism can also be applied in such a case \cite{Egger:2017tkd,Maas:2012tj,MPS:unpublished,Maas:2017swq}. Of course, there are now many more matrix elements on the right-hand side as in \prefr{fms1}{fms2}. Even for the simplest case at a lepton collider, i.\ e.\ two incoming, massless left-handed leptons described by composite operators $L$ and likewise two outgoing composite fermion operators $F$ \cite{Frohlich:1980gj,Frohlich:1981yi,Egger:2017tkd,Afferrante:2020fhd}
\bea
L&=&X^\dagger l=vl+\eta^\dagger l\label{ffms}\\
F&=&X^\dagger f=vf+\eta^\dagger f\nn,
\eea
\no where $l$ and $f$ are the elementary left-hand fermion fields, this becomes highly complicated\footnote{Note that likewise to the vector triplet \pref{wfms} this yields doublets of the global symmetry carried by the Higgs field $X$ \cite{Frohlich:1980gj,Frohlich:1981yi,Egger:2017tkd,Afferrante:2020fhd}. Thus, left-handed electrons and electron neutrinos in the standard model are really distinguished by the same quantum number as the physical $W$ and $Z$. The right-handed electron and electron neutrinos carry an independent right-handed flavor symmetry. The Yukawa couplings of the standard model eventually break both symmetries down to a common diagonal group \cite{Egger:2017tkd,Afferrante:2020fhd}. The same reasoning applies to quarks \cite{Maas:2017wzi,Egger:2017tkd}.}. Symbolically, in the center-of-mass frame \cite{Egger:2017tkd},
\be
\la \overline{L}(-p)L(p)\overline{F}(-q)F(q)\ra=v^4\la \overline{l}(-p)l(p)\overline{f}(-q)f(q)\ra+...\nn.
\ee
\no The remaining terms have less powers in $v$, but always one or more composite operator of type $(\eta^\dagger a)(k)$, where $a$ and $k$ can be $l$ and $p$ or $f$ and $q$, respectively. Taking only the leading term in the FMS mechanism recreates the usual perturbative results to all orders in the coupling constants \cite{Maas:2017wzi,Egger:2017tkd}, again confirming the reason for perturbation theory to work quantitatively well. 

Calculating the remaining terms requires to take the composite nature of the external fields into account. Such an augmented perturbation theory \cite{MPS:unpublished} works along the same lines as for matrix elements \cite{Maas:2020kda,Dudal:2020uwb}. In addition, it is necessary to supplement for the external states Bethe-Salpeter/Faddeev amplitudes \cite{Meissner:2022odx,MPS:unpublished}, instead of the non-interacting wave-functions of perturbation theory \cite{Bohm:2001yx}. They also need to be calculated consistently using the FMS mechanism \cite{MPS:unpublished}. Furthermore, the FMS mechanism introduces in the amputated, connected matrix elements an additional vertex, a bound-state splitting vertex \cite{Maas:2020kda,Dudal:2020uwb,MPS:unpublished}, which corresponds to a replacement of the composite states with elementary fields in the FMS mechanism \cite{Maas:2020kda,Dudal:2020uwb}. While these are only minor additions to the Feynman rules of perturbation theory to create an augmented perturbation theory, in practice this creates many more loop diagrams \cite{Maas:2020kda,Dudal:2020uwb,MPV:unpublished} for the additional 15 matrix elements involved. Thus, a full expression for any process is a formidable task, and remains still to be obtained.

\begin{figure}
 \includegraphics[width=\textwidth]{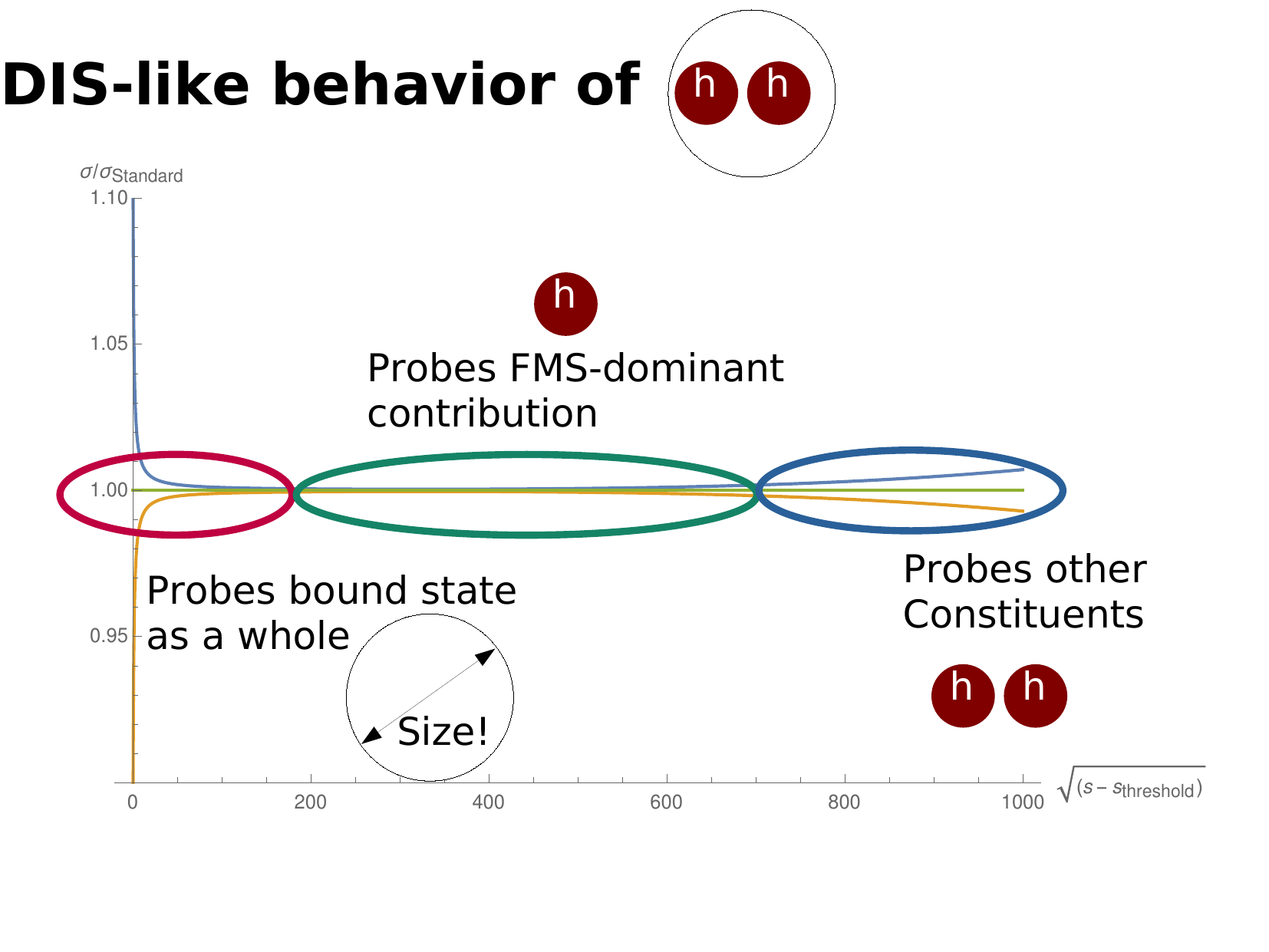}
 \vspace{-1.5cm}
 \caption{A cartoon sketch of the behavior of interactions between gauge-invariant composite states, given as the ratio to perturbative results as a function of an arbitrary energy value above threshold. For a bound state like $\det X=h^\dagger h$, at low energies the bound state as whole is probed. At intermediate energy, effectively only the first term in the expansion is relevant, giving the results closest to perturbation theory, as only the FMS-dominant contribution matters, here $h$. At very high energies the contribution of other valence particles and sea particles become probed.}
 \label{dis}
\end{figure}

However, there exist already a number of partial results \cite{Egger:2017tkd,Maas:2022gdb} as well as some lattice results \cite{Jenny:2022atm,Maas:2018ska}. They indicate that the interactions of the electroweak composite bound states will work in a similar vain as the interactions of hadrons in (deep) (in)elastic scattering (DIS). This is depicted schematically in figure \ref{dis}. There are three relevant energy regions, where energy here corresponds to some relevant energy scale $\sqrt{s}$. This could be, e.\ g., indeed the center-of-mass energy.

At very low energies, $m_\text{bound state}^2<s-N_\text{constituent}^2m_\text{constituent}^2\ll v^2$, the composite states are probed as a whole. This implies that their composite nature becomes readily apparent, e.\ g.\ the fact that they are not point-like \cite{Jenny:2022atm,Maas:2018ska}. Similar to the case of hadron physics, the probed extension depends on the involved probe particles \cite{BeiglboCk:2006lfa}. Since the extension is generated in the combination of weak and Higgs interactions, it likewise needs corresponding particles to probe it. While this information should be readily accessible from the Bethe-Salpeter/Faddeev amplitudes, these have not yet been calculated using augmented perturbation theory. However, lattice results indicate an effective size parameter of order a few inverse tens of GeV, both for the vector triplet and the scalar singlet \cite{Jenny:2022atm,Maas:2018ska}.

\begin{figure}
\begin{minipage}[c]{0.7\textwidth}
 \includegraphics[width=0.9\textwidth]{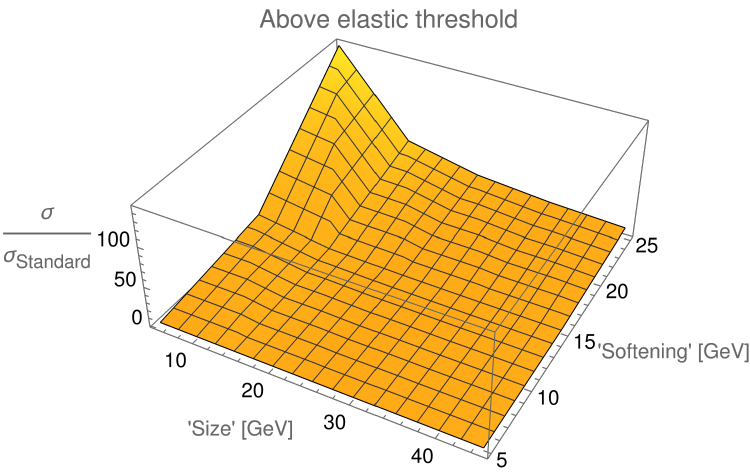}\\
 \includegraphics[width=0.9\textwidth]{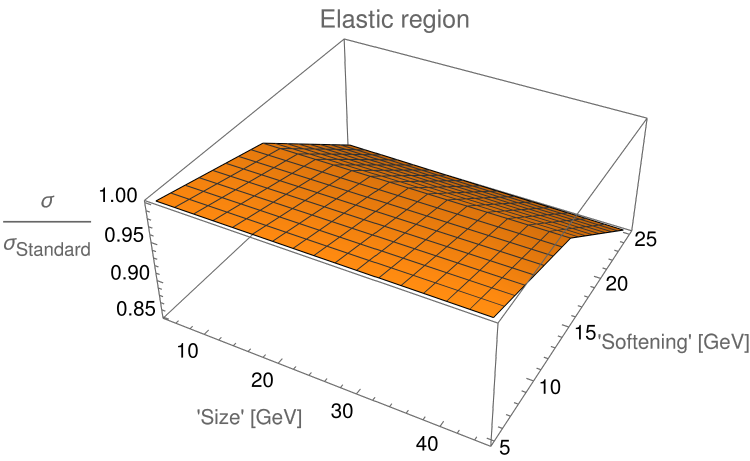}\\
 \includegraphics[width=0.9\textwidth]{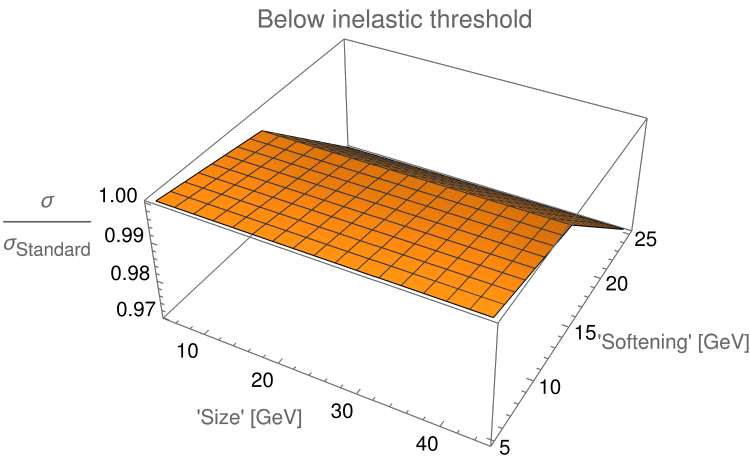}\\
 \end{minipage}\hfill
 \begin{minipage}[c]{0.3\textwidth}
 \caption{The deviation of the cross section with respect to the (Born-level) cross section for on-shell vector boson scattering \cite{Jenny:2022atm}. The three panels correspond to the integrated total cross section in three center of mass regions: Between 1 and 1.2 times the elastic threshold (top), between 1.2 and 1.5 times the elastic threshold (middle), and the remainder up to the inelastic threshold at 2 (bottom). The choice of range corresponds at low energies to that of ATLAS in \cite{ATLAS:2023dnm} and at intermediate energies to CMS in \cite{CMS:2022ley} for their control regions in the $ZZ\to4l$ final states. Size and softening correspond to the scattering length and the suppression of bound states effect at the first non-vanishing term in the threshold expansion, respectively, see \cite{Jenny:2022atm} for details.}
 \label{vbs}
 \end{minipage}\hfill
\end{figure}

Such an extension provides also a decisive test for the composite nature of the observed particles, and thus of the FMS mechanism. Vector boson scattering (VBS) \cite{Buarque:2021dji} appears as one suitable process to probe it \cite{Jenny:2022atm}. Here, for extensions of a few $1/10$ GeV$^{-1}$ as suggested by lattice simulations \cite{Maas:2018ska,Jenny:2022atm}, deviations within a few tens of GeV above the elastic threshold at $2m_V$ in the vector boson center of mass frame are expected, provided the vector bosons are on-shell, see figure \ref{vbs}.

This could be experimentally tested. Indeed, both the  ATLAS experiment \cite{ATLAS:2023dnm} and CMS experiment \cite{CMS:2022ley} at the LHC have shown that they can probe, in principle, the interesting regime. However, at the moment the contribution of VBS to the total yield is at most the same size as the uncertainties in the relevant kinematical regime \cite{ATLAS:2023dnm,CMS:2022ley}, and thus this awaits future improvements. At the same time, the theoretical estimates shown in figure \ref{vbs} are from a truncated standard model, and will also need to be pushed to quantitative precision for a final evaluation.

In the intermediate momentum regime with $s\sim v^2$, the states appear to be dominated entirely by the component which leads to the two-point function, e.\ g.\ the $W$ in \pref{wfms} or the lepton $l$ in \pref{ffms}. Thus, only a single valence particle dominates the state, the FMS-dominant constituent. Except for states like the scalar \pref{fms1} this is the non-Higgs valence particle. The Higgs valence particles act as spectators to this FMS-dominant constituent \cite{Jenny:2022atm,Maas:2020kda,Dudal:2020uwb,Maas:2018ska}. This is again very different from the QCD case, where all valence particles play qualitatively equal roles \cite{Dissertori:2003pj,BeiglboCk:2006lfa}. It is here where the results resemble closest the non-augmented perturbative ones. In particular, if the particle is produced as a resonance in this regime, this implies little to no change close to the pole \cite{Maas:2017wzi,Jenny:2022atm,Maas:2020kda,Dudal:2020uwb}.

At very high energies $s\gg v^2$, the other valence contributions \cite{Egger:2017tkd,Fernbach:2020tpa,Maas:2022gdb}, as well as sea contributions \cite{Bauer:2018xag}, start to become relevant. Depending on the colliding particles, this can have various impacts.

For colliding leptons, there will be additional contributions due to Higgs interactions \cite{Egger:2017tkd}. Especially, perturbative violations \cite{Ciafaloni:2000rp} of the Bloch-Nordsieck theorem \cite{Bloch:1937pw} will be canceled, as now full weak multiplets are present. This allows the treatment of collinear and soft radiation in the same way as in QCD \cite{Maas:2022gdb}. This deviation from non-augmented perturbative results increases like a double Sudakov-logarithm, i.\ e.\ roughly like $\ln^2 (s/m_W^2)$, at large energies. It can thus be a substantial effect at future lepton colliders \cite{EuropeanStrategyforParticlePhysicsPreparatoryGroup:2019qin}.

At the current LHC, where hadrons collide, consequences are less obvious due to the strong interaction background. Still, at the very least, changes in the parton structure of the hadrons will be needed \cite{Maas:2022gdb,Fernbach:2020tpa} as most hadrons, and especially protons, need a Higgs valence contribution for gauge invariance \cite{Maas:2017wzi,Egger:2017tkd,Maas:2022gdb}. These changes to the parton distribution functions may influence reactions involving particles coupling strongly to the Higgs, like top quarks and weak bosons \cite{Fernbach:2020tpa}. But due to the need to include this information in parton distribution function fits, this is far from being really testable \cite{Maas:2022gdb,Fernbach:2020tpa} yet.

\section{Applications beyond the standard model}\label{s:bsm}

The decisive feature for perturbation theory to work so well in the standard model is the one-to-one mapping of the gauge multiplet structure to the global multiplet structure of the global Higgs symmetry, e.\ g.\ in \pref{wfms} and \pref{ffms}. This similarly applies to all other standard model particles \cite{Frohlich:1980gj,Frohlich:1981yi,Maas:2017wzi,Egger:2017tkd}. In addition, there appear to be no additional bound states and resonances due to the weak bound state substructure, see \cite{Maas:2017wzi}. However, this may not be generally true \cite{Egger:2017tkd,Greensite:2020lmh}.

It is natural to ask \cite{Maas:2015gma} whether the same applies to general Yang-Mills-Higgs theories, with or without further matter. In the original works \cite{Frohlich:1980gj,Frohlich:1981yi}, it was already conjectured that adding further Higgs doublets to the standard model case would not alter the outcome qualitatively. Indeed, a detailed application of tree-level augmented perturbation theory to 2-Higgs doublet models \cite{Maas:2016qpu} and the minimal supersymmetric standard model \cite{Schreiner:2022ms,Maas:2023ms} confirm this conjecture. A check using non-perturbative methods has, however, not yet been performed. But it would be feasible to do so, e.\ g.\ using lattice methods \cite{Wurtz:2009gf,Lewis:2010ps,Sarkar:2022rti}. Such models enlarge the global group, while leaving the gauge group fixed. However, because any phenomenological viable version of these theories supports global SU(2) multiplets \cite{Branco:2011iw,Aitchison:2007fn}, the underlying FMS mechanism still works \cite{Maas:2016qpu,Maas:2023ms}.

The situation is potentially very different when enlarging the gauge group, while keeping or reducing the global group \cite{Maas:2017wzi,Maas:2015gma}. This is the situation typical for grand unified theories (GUTs) \cite{Langacker:1980js,O'Raifeartaigh:1986vq}. There are two separate aspects to be taken care of \cite{Maas:2017xzh}. One is that such theories offer usually multiple breaking patterns in the BEH effect instead of the single one in the standard model. The other is that the accompanying change of the global group with respect to the standard model alters the possible multiplet structure of observable, gauge-invariant states.

Because the FMS mechanism is a map between the gauge-invariant operators and gauge-dependent operators, it needs a prescription to which gauge states it should map, a choice of gauge. If different breaking patterns can be realized by different gauge choices, this yields different possibilities for the map, and thus potentially different results of the FMS mechanism for the physical spectrum. This would jeopardize the usability of the FMS mechanism \cite{Maas:2017xzh}.

The simplest example for which this happens is an SU(3) gauge group with a single Higgs in the adjoint representation \cite{O'Raifeartaigh:1986vq}. At tree-level, two equivalent breaking patterns occur, SU(2)$\times$U(1) and U(1)$\times$U(1), which differ, e.\ g.\, in the number of massless gauge bosons \cite{Dobson:2022ngz,O'Raifeartaigh:1986vq,Maas:2017xzh}. In such a case multiple possibilities exist how to set up the FMS mechanism \cite{Maas:2017xzh}, which will lead to different, and inequivalent, predictions for the physical spectrum. However, in general already one-loop corrections will lift the tree-level degeneracy \cite{Dobson:2022ngz,Kajantie:1998yc,O'Raifeartaigh:1986vq}. Thus, there is again only one gauge choice in which the field fluctuations can be considered small, yielding again a unique setup for the FMS mechanism \cite{Dobson:2022ngz,Maas:2017xzh,Sondenheimer:2019idq}. Thus, this appears to be not an issue, as long as no exception is found. However, this also implies that to apply augmented perturbation theory in such cases requires to first determine the corresponding allowed breaking pattern at the same order for the parameter set in question. This will be assumed to have happened in the following.

The physical spectrum can only form representations of the global symmetry groups. If these do not support the same multiplicities as the unbroken gauge groups, it cannot be expected that the one-to-one mapping of the standard model still works \cite{Maas:2015gma}. This is indeed the case.

The simplest example is a SU(3) Yang-Mills theory coupled to a Higgs field $\phi$ in the fundamental representation. The BEH effect creates the pattern SU(3)$\to$SU(2), as the only possible little group in this case. Thus, this yields 3 massless gauge bosons in the adjoint of the unbroken SU(2), as well as 5 massive ones, of which four are degenerate and form two doublets under the unbroken SU(2), and one singlet under the unbroken SU(2) \cite{Bohm:2001yx,O'Raifeartaigh:1986vq,Maas:2017xzh}. At the same time, the global symmetry is merely a U(1) acting as a global phase on the Higgs field \cite{Maas:2016ngo}. Hence, in absence of accidental degeneracies, it can at most support two particles of the same mass. They form a particle and an antiparticle with respect to the U(1) group.

Augmented perturbation theory is in agreement with this analysis. The decisive ingredient is the matrix $c$ from \pref{wfms}. The simplest operator to understand the difference creates a U(1)-neutral vector state. Using the FMS mechanism it follows that \cite{Maas:2016ngo}
\be
\phi_i D_\mu^{ij} \phi_j=v^2c_a W_\mu^a+...\nn,
\ee
\no where the mapping 'matrix' $c$ now has the form $c_a=\delta_{a8}$, if the Higgs vacuum expectation value has been given the real 3-direction. In this way, the FMS mechanism maps the physical state to the most massive gauge boson, the singlet under the unbroken SU(2) gauge subgroup. This pattern continues to other channels \cite{Maas:2017xzh}. Especially, as no symmetry exists to create a gauge-invariant triplet, the prediction from augmented perturbation theory is that the theory is gapped \cite{Maas:2017xzh,Maas:2016ngo,Maas:2018xxu}, in stark contrast to the gauge-dependent, ungapped spectrum. Again, augmented perturbation theory has been confirmed for this theory in lattice simulations \cite{Maas:2016ngo,Maas:2018xxu,Dobson:2023,Dobson:2022crf}.

\begin{figure}
\begin{minipage}[c]{0.6\textwidth}
 \includegraphics[width=0.9\textwidth]{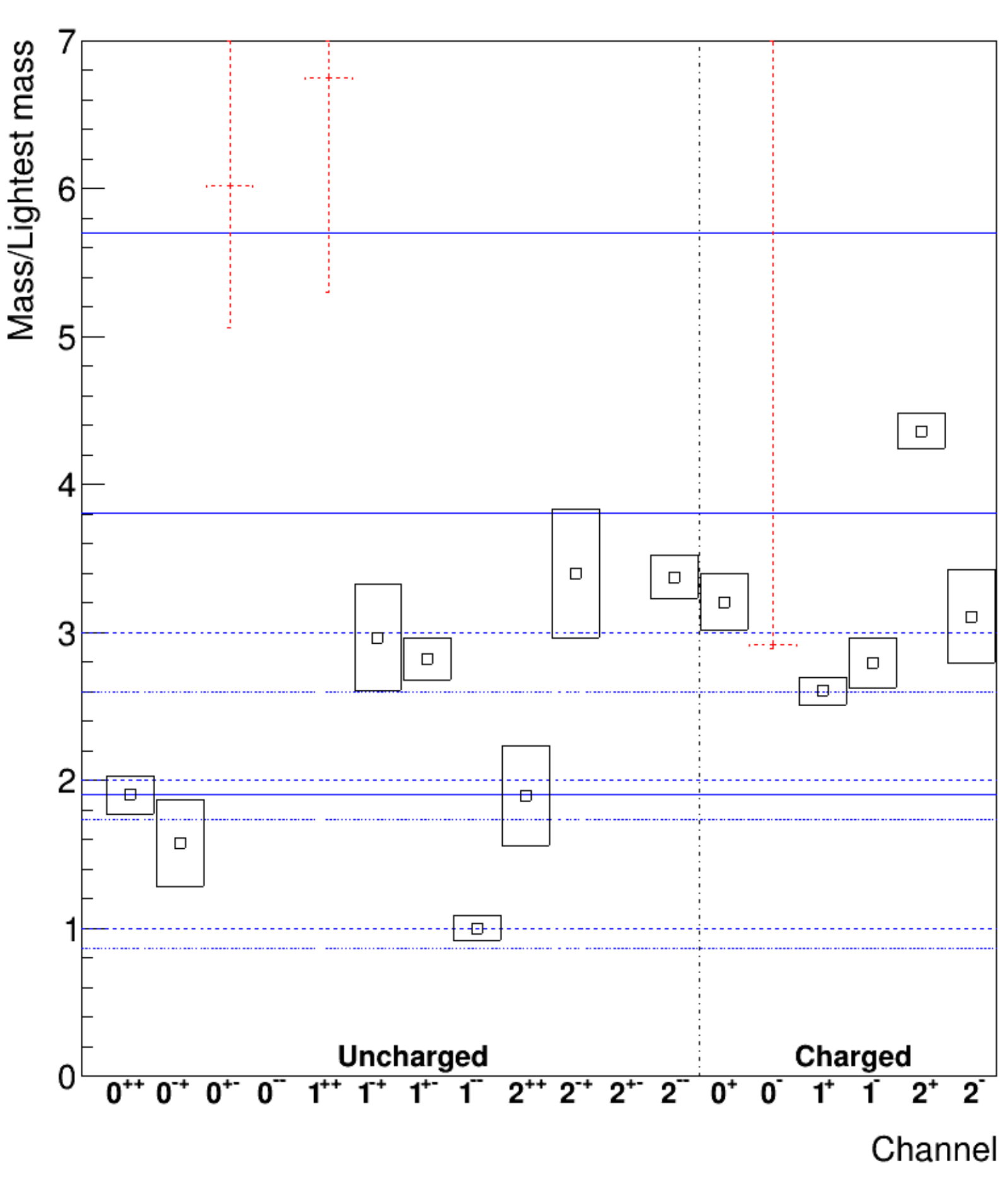}
 \end{minipage}\hfill
 \begin{minipage}[c]{0.4\textwidth}
 \caption{The lowest level of SU(3) with a fundamental Higgs in the BEH Higgs region in various quantum number channels in units of the lightest state using lattice methods. The results are infinite-volume extrapolated and used a variational analysis. Simulations have been performed at $\beta=6.85535$, $\kappa=0.456074$, and $\lambda=2.3416$, see \cite{Maas:2018xxu,Dobson:2023,Dobson:2022crf} for further technical details. Results, partially preliminary, are from \cite{Dobson:2023,Dobson:2022crf}. Red dashed lines indicate upper limits and the horizontal blue lines correspond to the gauge-dependent mass scales, with the solid line the elementary Higgs mass, the dashed line the heavy gauge boson mass, and the dotted line the lighter gauge bosons´ mass, as well as integer multiples. ``(Un)Charged'' refers to the U(1) charge and the label to the $J^{PC}$ and $J^P$ quantum numbers for the charged and uncharged states, respectively.}
 \label{specbsm}
 \end{minipage}\hfill
\end{figure}

Moreover, similar to QCD, it is not possible to construct a gauge-invariant state with only a single unit of U(1) charge, but at least 3 units are required. This is again in stark contrast to the perturbative case. Such states should perturbatively not appear asymptotically in a different way than as a scattering state. But non-perturbatively, because the U(1) charge is conserved, at least the lightest state with 3 units of U(1) (as well as its antistate) needs to be necessarily stable. Such a state is indeed observed in lattice simulations \cite{Maas:2018xxu,Dobson:2022crf}. However, because there is no charge-3 elementary state, its gauge-invariant operator cannot be mapped by the FMS mechanism onto an elementary state. Thus, the first non-zero matrix element in the FMS expression is not a propagator, but a higher $n$-point function \cite{Maas:2017xzh}. At lowest order in a constituent-like model, this matrix element seems to provide at least the correct order of magnitude of the mass \cite{Maas:2017xzh,Maas:2018xxu,Dobson:2022crf}, but a detailed investigation is yet required. In total, the spectrum becomes indeed very involved and rich, once more channels are taken into account \cite{Dobson:2023,Dobson:2022crf}, see figure \ref{specbsm}.

Going beyond this simplest example, it becomes quickly apparent that one big difference is the number of physical degrees of freedom in contrast to the gauge degrees of freedom. Consider an SU($N$) gauge theory coupled to a single fundamental Higgs. While the former is determined, up to internal excitations, by the possible multiplet structure of the global symmetry, which is fixed to U(1), the latter quickly increases with increasing dimension of the gauge group \cite{Maas:2017xzh,Sondenheimer:2019idq}. Thus, the low-energy physics remains very similar, just as is the case with the glueball physics of large-$N$ Yang-Mills theory. Also there the number of physical states remains fixed, and essentially unaltered, even when moving towards an infinite number of gauge degrees of freedom \cite{Lucini:2012gg}. Thus, such a behavior is not without precedent.

Going beyond this case,  many aspects of the spectrum become quickly dependent on the details of the representations of the Higgs fields and breaking patterns \cite{Maas:2017xzh,Sondenheimer:2019idq}. But two more general features stand out.

The first is that any GUT \cite{Langacker:1980js,O'Raifeartaigh:1986vq} needs to also include electromagnetism and thus the photon. The photon is exceptional, as it is a physically observable, massless vector state. This is well understood, but non-trivial, in QED \cite{Haag:1992hx,Lenz:1994tc}. But in a GUT setting the requirement of gauge invariance requires that the photon is also a composite state, which is gauge-invariant with respect to the single unified gauge group. Hence, this requires that there exists a massless, uncharged composite vector bound state. This is indeed predicted with augmented perturbation theory \cite{Maas:2017xzh,Sondenheimer:2019idq}, but requires as minimal Higgs content a Higgs in the adjoint representation. While the calculation is involved, the appearance of such a state was confirmed in lattice gauge theory at an exploratory level \cite{Kajantie:1998yc,Lee:1985yi,Afferrante:2020hqe}. This also provides an explicit example for the possibility to build a massless vector boson from massive constituents, without involving a Goldstone-type mechanism. In particular, given the explicit mass scale in the Lagrangian, even the usual argument of interpreting the photon as the Goldstone mode of broken dilatation symmetry \cite{Lenz:1994tc} does not apply here.

The second feature is the low-energy behavior. If the ideas of GUTs should work, it is required that the physical, gauge-invariant low-energy spectrum of the standard model is reproduced. While this is well reproduced in a perturbative approach \cite{Langacker:1980js}, this is not true for the gauge-invariant spectrum \cite{Maas:2017xzh,Sondenheimer:2019idq}. Here, the pattern of the simplest example repeats itself. Especially, many popular GUT candidates with minimal Higgs content are explicitly ruled out on a qualitative mismatch of the spectrum \cite{Sondenheimer:2019idq}. In fact, no qualitatively working candidate has yet been found \cite{Maas:2017xzh,Sondenheimer:2019idq}, not to mention obeying quantitative constraints like proton decay \cite{Bohm:2001yx,Langacker:1980js}. Whether it is possible at all is a difficult question, and it could be even impossible in the conventional way \cite{Sondenheimer:2019idq}. At the very least it appears not possible without having a Higgs content able to completely break the gauge group, while at the same time having a suitable global symmetry structure \cite{Sondenheimer:2019idq}.

\begin{figure}
\begin{minipage}[c]{0.75\textwidth}
 \includegraphics[width=\textwidth]{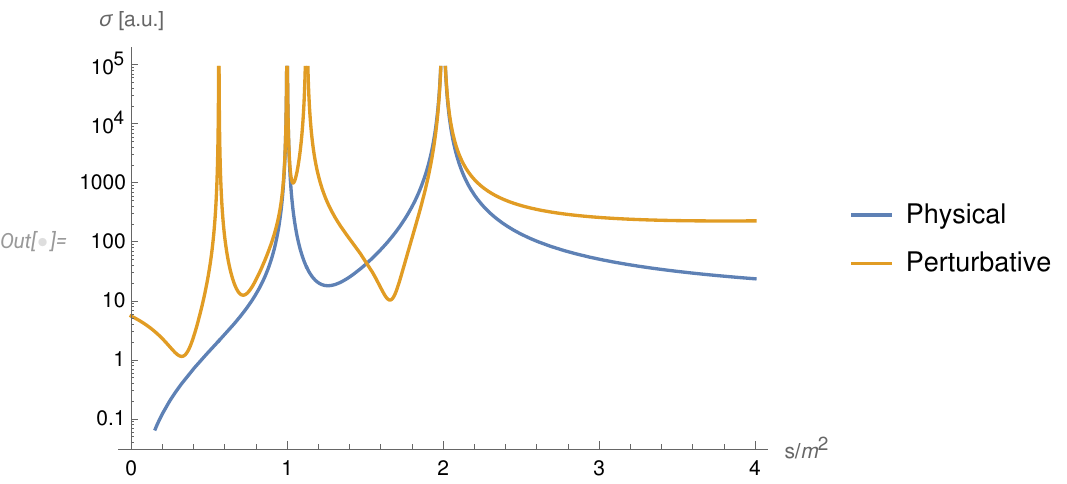}
 \end{minipage}\hfill
 \begin{minipage}[c]{0.25\textwidth}
 \caption{The tree-level scattering cross-section in arbitrary units for Bhabba scattering at zero rapidity in units of the singlet $1^{--}$ mass in figure \ref{specbsm} for the theory described in the text \cite{Maas:2017wzi}. Perturbative results are given in gold, and augmented perturbative results in blue.
 }\label{scattbsm}
 \end{minipage}\hfill
\end{figure}

There are further far-reaching implications for searches for new physics because of the different spectrum. Consider again as the simplest example an SU(3) gauge theory with a Higgs in the fundamental representation and coupled in addition to a fermion in the fundamental representation coupled vectorially to the gauge interaction. This allows for an explicit mass, and does not require a Yukawa coupling to the Higgs. Performing Bhabba scattering it would usually be assumed that in the $s$-channel in the intermediate states all gauge bosons would show up, already at tree-level \cite{Bohm:2001yx,Maas:2017wzi}. However, treating the fermions correctly as bound states changes this qualitatively \cite{Maas:2017wzi,Torek:2018qet}. The fermion bound state does no longer couple to all of the gauge bosons at tree-level, as the corresponding matrix element of bound states now contains a Higgs vacuum expectation value, which projects out all couplings, which are not mapped by the FMS mechanism to a vector bound state. Thus, only those vector states show up as resonances, which also appear in the physical spectrum. This is shown in figure \ref{scattbsm}. It is very satisfying that the resonances in the cross section are indeed only the physical ones. However, if experimental searches would not be based on augmented perturbation theory, but on perturbation theory, such a theory could be wrongly excluded. Especially if the energy reach is too small, where the gapped nature of the physical spectrum makes itself felt. Besides that, the fake resonances coming from unphysical degrees of freedom could easily misguide experimental searches. This does need to be taken properly into account. It should also be noted that close to the physical $s$-channel resonance (and the corresponding $t$-channel resonance) the results are nonetheless essentially indistinguishable. This had also been observed at next-to-leading order for two-point matrix elements \cite{Maas:2020kda,Dudal:2020uwb}.

\section{Applications beyond Yang-Mills-Higgs theory}\label{s:bym}

While the FMS mechanism proper requires a BEH effect to work as described, the FMS framework can be extended far beyond. This potential has almost not been tapped at all yet.

\subsection{Theories without a Higgs}

The FMS mechanism requires something to expand around. This is the reason, why it does not work, e.\ g., in QCD\footnote{Though similar ideas can be pursued \cite{Hoyer:2016orc}.}. But the FMS framework still requires to take manifest gauge invariance into account, and always start from there.

To understand the implications, consider the idea that the Higgs is a composite state of new fermions \cite{Lane:2002wv,Sannino:2009za,Andersen:2011yj}, either in the context of technicolor or some other scenario. While the additional sector is strongly interacting, the weak interactions remains what it is. Thus, there are still weak gauge bosons, which need to be dressed to obtain the observable gauge-invariant (almost degenerate) vector state triplet \pref{wfms}, the physical version of the $W$-bosons and the $Z$ boson \cite{Maas:2015gma}.

If foregoing the possibility to construct a low-energy effective theory \cite{Andersen:2011yj} and rather work with the ultraviolet theory, this immediately shows a problem. At least some of the additional fermions $\psi$ need to be charged under both, the weak interactions and the new interactions. The former, because otherwise the new sector decouples. The latter, because otherwise the Higgs cannot be a bound state.

While the Higgs can be constructed as a straightforward singlet, the situation becomes involved for the vector bosons. The simplest possibility is to have two non-degenerate fermions, and to build an operator like
\be
\tau^a_{ij}\bar{\psi}^i_{ru}\gamma^\mu D_\mu^{ru;sv}\bar{\psi}^i_{ru}\label{techniw}
\ee
\no where $aij$ are flavor indices, $rs$ are new interaction indices, and $uv$ are weak indices. The covariant derivative therefore contains both the weak gauge bosons and the new gauge bosons. This creates a state similar to the $\rho$ meson of QCD, where the mass-splitting of the fermions can create the difference in the $W$ and $Z$ masses, and the SU(2) generator counts with $a$ these states.

While this is formally working, this leads to two problems. On the one hand, phenomenology tends to require rather a large number of flavors \cite{Sannino:2009za,Andersen:2011yj}. Their mass splitting must therefore be substantial to the lightest doublet, to avoid creating further vector states. At the same time, this is at odds with the foundational principle \cite{Lane:2002wv,Sannino:2009za,Andersen:2011yj} of using a scaled-up version of QCD \cite{Quigg:2009xr} to create masses by a condensate. In this scenario the pseudo-Goldstones would be absorbed as longitudinal degrees of freedom from the vector states, to avoid having light pseudoscalars around. With an operator like \pref{techniw} this is neither possible, nor necessary. Also, such Goldstone bosons could not be defined in a gauge-invariant way with respect to the weak interactions, if they should play this role. Thus, such a scenario appears to be inconsistent with gauge symmetry. Moreover, quantitatively it would be required that a state created by \pref{techniw} is the lightest (visible) state in the spectrum of the new sector. Whether this is possible, is unknown. At least, no working example has been found yet \cite{Maas:2017wzi}. But essentially no dynamical investigations respecting manifest gauge symmetry with both sectors coupled have been conducted yet either.

Thus, also here the FMS framework shifts qualitatively the way how scenarios of this type needs to be addressed. Perhaps this will also open quite different possibilities, as this alleviates some of the problems due to light states in such theories.

\subsection{Gravity and supergravity}

\subsubsection{Gravity}

The FMS mechanism relies on the single-field expansion \pref{behsplit}. Therefore, only scalar fields can be used for the expansion, as long as global Poincar\'e invariance should be maintained. This is true in all quantum field theories. Hence, in many theories the FMS mechanism is not applicable. There is, however, a further exciting possibility beyond quantum field theory: Gravity.

General relativity can be considered a gauge theory of translations \cite{Hehl:1976kj,Kibble:1961ba,Sciama:1962}. Local Lorentz symmetry in the tangent space can then be considered to be either dependent on the translation gauge symmetry, or can be considered as a second gauge interaction, leading to torsion \cite{Hehl:1976kj}. Correspondingly, a quantum gravity theory will also be a quantum gauge theory. It is yet unclear if a, more or less extended, version of canonical gravity can be quantized using a path-integral, or similar, approach. But there is encouraging mounting circumstantial evidence that this is the case \cite{Ambjorn:2012jv,Loll:2019rdj,Niedermaier:2006wt,Reuter:2012id,Reuter:2019byg,Christiansen:2015rva,Bosma:2019aiu,Schaden:2015wia,Asaduzzaman:2019mtx,Catterall:2009nz}. It will therefore be assumed here. The FMS framework then applies as well, requiring to start out with manifestly diffeomorphism-invariant (and local Lorentz-invariant) quantities as observables \cite{Ambjorn:2012jv,Loll:2019rdj,Maas:2019eux}

Observationally, at long distances, quantum gravity is indeed dominated by a special field configuration in our universe, a de Sitter metric\footnote{On the question whether this should rather be a Friedmann-Lem\^aitre-Robertson-Walker metric see \cite{Maas:2022lxv}.}. This suggests that it should at least be possible to apply also the FMS mechanism in this case \cite{Maas:2019eux,Maas:2022lxv}. Conceptually, this creates a BEH effect in quantum gravity. Similarly to the BEH effect in quantum field theory, the classical minimum will be the starting point. For the observational value of Newton's constant and the cosmological constant \cite{pdg}, this is indeed de Sitter. This is in agreement with the metric structure at long distances, supporting the possibility that the FMS mechanism should be possible.

There is one particularity which makes it different from the situation so far. In the previous cases, the field developing the vacuum expectation value and the gauge fields were not the same. Here, they are. While this does not introduce any conceptual obstacles, it makes it technically more involved. This is amplified by involvement of the gauge field, the metric, itself in all expressions.

Probably the most cumbersome feature is that a BEH effect in quantum gravity requires necessarily the use of a non-linear gauge condition. The FMS mechanism can only be applied after gauge-fixing and quantization is complete, and therefore does not alter it.

Consider as a minimal case the Einstein-Hilbert action
\bea
S&=&\frac{1}{2\kappa}\int d^4x\sqrt{\det(-g)}\left(R+l\right)\nn\\
\Gamma_{\mn\rho}&=&\frac{1}{2}\left(\pd_\rho g_\mn+\pd_\nu g_{\mu\rho}-\pd_\mu r_{\nu\rho}\right)\nn\\
R&=&g^\mn R_\mn=g^\mn g^{\rho\sigma} R_{\rho\mu\sigma\nu}\nn\\
R_{\mu\nu\rho\sigma}&=&\pd_\rho\Gamma_{\mn\sigma}-\pd_\sigma\Gamma_{\mn\rho}+g^{\alpha\beta}\Gamma_{\mu\alpha\rho}\Gamma_{\beta\nu\sigma}-g^{\alpha\beta}\Gamma_{\mu\alpha\sigma}\Gamma_{\beta\nu\rho}\nn
\eea
\no where $g_\mn$ is the metric, $R$ the curvature scalar, and $\kappa$ and $l$ suitably normalized versions of Newton's constant and the cosmological constant, respectively.

There is now a choice to be made, which was not previously present: Should the gauge condition be obeyed by the field configuration to be expanded around in the FMS mechanism? Before that, any choice here could be compensated by the gauge field. But now the vacuum expectation field and the gauge field is identical. It appears to be technically convenient \cite{Maas:2022lxv} to choose a gauge, which is satisfied by the vacuum expectation value, in the present case the de Sitter metric $g_\mn^\text{dS}$. Due to reparametrization invariance, this can furthermore substantially alter the technical feasibility. One possible practical choice appears to be the Haywood gauge \cite{Maas:2022lxv},
\be
g^\mn\pd_\mu g_{\mu\rho}=0\nn,
\ee
\no which is fulfilled by a flat metric, and the maximal symmetric de Sitter and anti-de Sitter metrics in standard Cartesian parametrization. This condition already shows the issue of non-linearity. In particular, because the inverse metric is not independent. This relation is in general highly non-linear.

The FMS mechanism constitutes now in splitting off the ``vacuum expectation value '' $g_\mn^\text{dS}$ \cite{Maas:2019eux}. Again, there is no unique way to do so. But if any approximations are good, the split-off part $\gamma_\mn$ needs to be small, and thus at linear order many possibilities coincide \cite{Maas:2022lxv}, yielding
\be
g_\mn=g_\mn^\text{dS}+\gamma_\mn\nn.
\ee
\no The inverse of $\gamma$ is not a metric, and determined by a Dyson-like relation
\be
\gamma^\mn=-(g^\text{dS})^{\mu\sigma}\gamma_{\sigma\rho}g^{\rho\nu}\label{dyson}.
\ee
\no Because $\gamma_\mn$ is assumed small, the right-hand side can be expanded in a series in $\gamma_\mn$. This creates an infinite series of tree-level vertices \cite{Maas:2022lxv}, but establishes a formulation in $\gamma_\mn$ only. And a similar step is necessary for most observables as well.

While this is technically more cumbersome than in the quantum-field theory case, it is straightforward to use \cite{Maas:2019eux,Maas:2022lxv}. At tree-level, it yields agreement with results from dynamical triangulation \cite{Ambjorn:2012jv,Loll:2019rdj,Maas:2022lxv,Dai:2021fqb} as well as a well-defined systematic limit to flat-space quantum-field theory as the lowest order in the FMS mechanism \cite{Maas:2019eux,Maas:2022lxv}.

The latter is probably best seen by considering how distances are measured in quantum gravity. Distance itself becomes in quantum gravity an expectation value \cite{Ambjorn:2012jv,Schaden:2015wia,Maas:2019eux}. A possible definition is given by
\bea
r(x,y)&=&\la \min_{z(t)}\int_{x}^{y} dt g^\mn \frac{dz_\mu(t)}{dt}\frac{dz_\nu(t)}{dt}\ra\nn\\
&=&\min_{z(t)}\int_{x}^{y} dt g_\mn^\text{dS} \frac{dz^\mu(t)}{dt}\frac{dz^\nu(t)}{dt}+\la \min_{z(t)}\int_{x}^{y} dt \gamma_\mn \frac{dz^\mu(t)}{dt}\frac{dz^\nu(t)}{dt}\ra,\nn
\eea
\no where $x$ and $y$ are points in the $\rr^4$ underlying the manifold and the minimization requires to find the geodesic distance\footnote{If singularities appear, or geodesics become incomplete, a suitable deformation has to be introduced.} between these points in the manifold configuration. This is averaged over the manifold configurations. The second line implements the FMS mechanism, which shows how the result splits between the contribution from the vacuum expectation value and the fluctuation field. Especially, if the fluctuations vanish, $\gamma\to 0$, this smoothly changes into the ordinary fixed curved-background quantum field theory distances.

\subsubsection{Supergravity}

Following the FMS framework through often leads to very surprising insights. Consider the concept of supersymmetry \cite{Aitchison:2007fn,Weinberg:2000cr}. In quantum field theory, supersymmetry appears to be essentially transparent for the FMS mechanism \cite{Maas:2023ms}. This is not surprising, as supersymmetry is a global symmetry, and thus should behave like, e.\ g., flavor symmetries.

However, despite all efforts and its inherently appealing nature \cite{Aitchison:2007fn,Weinberg:2000cr}, no sign of supersymmetry in nature has been observed \cite{pdg}. This leads to the claim that supersymmetry must be necessarily broken, at the expense of loosing some of its appeal \cite{Aitchison:2007fn,Weinberg:2000cr}.

It is here were the FMS framework provides a possible way out. In our actual universe, it is not valid to consider supersymmetry as a stand-alone global symmetry, due to the existence of gravity. Because supersymmetry is part of the super Poincar\'e symmetry, this forces supersymmetry to become a local gauge symmetry, supergravity \cite{VanNieuwenhuizen:1981ae,Freedman:2012zz}. According to the FMS framework, physical observables cannot be gauge-dependent, and cannot change under the local supersymmetry transformations. Thus, the physical, observable spectrum is not and, in fact, cannot be supersymmetric. This alleviates the need to find a superpartner for, e.\ g., the electron, which is the usual argument for requiring supersymmetry to be broken \cite{Aitchison:2007fn}. Thus, it is possible to retain supersymmetry, and supergravity, as an intact symmetry of nature, without the need to observe a supersymmetric spectrum at experiments. Given the importance of supersymmetry to string theory \cite{Polchinski:1998rr}, this can have far-reaching consequences.

In addition, similar to canonical quantum gravity, this implies the possibility to use the FMS mechanism on the same reasoning, this time introducing a BEH effect for the vierbein $e_\mu^a$. Consider the simplest ${\cal N}=1$ supergravity theory \cite{Weinberg:2000cr}
\bea
S&=&\int dx\frac{\det e}{2\kappa}\left(e^{a\mu}e^{b\nu}R_{\mn ab}-\bar{\Psi}_\mu\gamma^{\mn\rho}D_\nu\Psi_\rho\right)\nn\\
R_{\mn ab}&=&\pdm\omega_{\nu ab}-\pdn\omega_{\mu ab}+\omega_{\mu ac}\omega_{\nu\,\, b}^{\,\,c}-\omega_{\nu ac}\omega_{\mu\,\,b}^{\,\,c}\nn\\
D_\nu&=&\pd_\nu+\frac{1}{4}\omega_{\nu ab}\gamma^{ab}\nn\\
\omega_{\nu ab}&=&2e_{\mu[a}\pd_{[\nu}e_{b]}^{\mu]}-e_{\mu[a}e_{b]}^\sigma e_{\nu c}\pd^\mu e_\sigma^c\nn\\
g_\mn&=&e_\mu^a g^\text{Flat}_{ab} e^b_\nu\nn,
\eea
\no with the Rarita-Schwinger graviton $\Psi$ and $[i,j]$ implies antisymmetrization of $i$ and $j$. Under a supersymmetry transformation $\delta_S$
\bea
\delta_S e_\mu^a&=&\frac{1}{2}\bar{\epsilon}\gamma^a\Psi_\mu\nn\\
\delta_S\Psi_\mu&=&D_\mu\epsilon\nn,
\eea
\no with local transformation function $\epsilon(x)$. As this is a gauge transformation, any physical observable needs to be invariant under it.

A possible example is the local composite operator\footnote{Torsion will require a similar treatment for the $\gamma$ matrices, probably using position-dependent $\gamma$ matrices \cite{Gies:2013noa,Gies:2015cka} and another FMS mechanism for them \cite{Maas:2019eux}.}
\be
e_a^\mu\gamma^a\Psi_\mu\nn.
\ee
\no It is invariant under a supersymmetry transformation due to the compatibility of the tetrad and the Grassmann nature of the graviton. It therefore does not have a superpartner.

At the same time, applying the FMS mechanism in Haywood gauge with $e_\mu^a=\delta_\mu^a+\varepsilon_\mu^a$ for flat space and a small fluctuation field $\varepsilon$ yields
\be
e_a^\mu\gamma^a\Psi_\mu=\gamma^\mu \Psi_\mu+\varepsilon^a_\mu\gamma^a \Psi_\mu\nn.
\ee
\no Neglecting, as usual, the fluctuation part, the state describes a massless spin 1/2 particle. While promising to be more involved than the ordinary gravity case \cite{Maas:2022lxv}, it appears very appealing to follow-up on these exploratory heuristics.

\section{Ontological implications}\label{s:phil}

The FMS framework is from the point of view of philosophy of physics quite remarkable from two perspectives \cite{Berghofer:2021ufy,Friederich:2013,Friederich:2014,Lyre:2004,Lyre:2008af,Francois:2018,Smeenk:2006,Struyve:2011nz}. One is from the resolution of ambiguities in the BEH effect using the FMS mechanism. The other concerns the implications for the laws of nature by the FMS framework.

The FMS mechanism resolves the apparent paradox \cite{Lyre:2004,Lyre:2008af,Struyve:2011nz} of the gauge-dependence of the BEH effect and its apparent phenomenological success when treated as if it would be physical \cite{Maas:2017wzi,Maas:2012ct,Lee:1974zg,Caudy:2007sf}. It shows that the paradox is an artifact of the special structure of the standard model \cite{Maas:2017wzi}, which allows for a quantitatively effective possibility to ignore the issue of non-perturbative gauge invariance. Still, it was, and is, a source of some consternation in the philosophy of physics literature \cite{Berghofer:2021ufy,Lyre:2008af,Francois:2018,Struyve:2011nz}, why this paradox has not been taken seriously, not realized, or even denied in large parts of the particle physics community. In fact, this dissonance even led to the odd situation that lattice approaches, which need by construction to take non-perturbative gauge invariance seriously and manifestly into account, denoted the composite states Higgs, $Z$, $W$, and so on \cite{Lang:pc,Wittig:pc,Evertz:1986ur,Langguth:1985dr,Philipsen:1996af}, in obvious contradiction to their nature. Hence, despite having with the FMS mechanism a conceptually clean approach, the underestimation of the mechanism leaves still a kind of quagmire in notations in contemporary literature. Philosophically, of course, posing the question what is real, and what the role of gauge symmetry is, leads immediately to the necessity to find a resolution of the paradox.

This leads to the even more important perspective, this time with respect to the FMS framework. The question for the role of gauge symmetries is a very fundamental one. Since it appears possible to remove them from some theories explicitly \cite{Frohlich:1980gj,Frohlich:1981yi,Philipsen:1996af,Langguth:1985eu,Masson:2010vx,Attard:2017sdn,Kondo:2018qus,Gambini:1996ik,Strocchi:1974xh}, it is questionable whether they have any ontological relevance at all. This has already been formulated in terms of the Kretschmann objection \cite{Pooley:2017}, which in its generalized form states that any theory can be turned into a gauge theory, and in its inverse form that every gauge theory can be rewritten in terms of a (possibly non-local) non-gauge theory \cite{Berghofer:2021ufy}. It is a most remarkable feature that such non-gauge theories seem at first sight to be again a theory of point-particles. However, due to the appearance of either an infinite series of polynomials in the Lagrangian and/or non-localities, it becomes quickly evident that this is just an artifact of tree-level perturbation theory \cite{Maas:2017wzi}. In this context, it is important to note that the Aharanov-Bohm effect \cite{Aharonov:1959fk}, often cited as supporting that gauge fields are physically real, can indeed also be described entirely without resorting to gauge degrees of freedom \cite{Berghofer:2021ufy,Strocchi:1974xh}.

It thus appears that gauge symmetries are merely redundant degrees of freedom\footnote{It has been claimed, see e.\ g.\ \cite{Reece:2023czb} for an introduction, that semi-classical considerations of black holes make gauge symmetries physical. It could not yet been substantiated whether this holds true in full quantum gravity \cite{Perry:pc}, and gauge-invariant formulations of canonical quantum gravity appear to disfavor such a possibility \cite{Ambjorn:2012jv,Loll:2019rdj,Maas:2019eux,Donnelly:2015hta,Giddings:2019wmj}.}, which are however technically indispensable. However, this is not a very precise phrasing, see \cite{Berghofer:2021ufy} for a more detailed discussion.

A possible stance is that only measurable, and thereby at least gauge-invariant, entities should be ontological, i.\ e.\ possible candidates for being part of reality. If one accepts  this premise, the FMS framework fundamentally reshuffles the building blocks of nature. Aside from hypothetical right-handed neutrinos, all observed particles are necessarily extended, and described by composite, gauge-invariant operators. This is a fundamental paradigmatic shift compared to the idea of fundamental point particles. It was also the latter idea, which gave rise to string theory \cite{Polchinski:1998rq}, due to the problems entailed by the point-like nature of elementary particles. Having as fundamental entities composite ones would change this premise at least partly.

Furthermore, allowing the fundamental laws of nature to be build from extended objects would possibly open up alternatives to the idea of ever smaller structures, or higher energies. Especially, as the concept of energy itself becomes in quantum gravity ontologically doubtful, as energy is no longer gauge-invariant. Such a recast of the approach to the fundamental laws of nature would be nothing but transformative, and would even affect school textbooks and popular science fundamentally.

\section{Summary}

The most obvious consequence of the FMS framework \cite{Frohlich:1980gj,Frohlich:1981yi}, and with this one aspect of Giovanni Morchio's legacy, is to reconcile the foundations of field theory with the phenomenological success of the perturbative treatment of the BEH effect. With the FMS mechanism, this delivered a tool to turn the very fundamental considerations of the FMS framework into phenomenological applications and even paved the way to experimental tests. In fact, it will allow a guaranteed discovery. Either, experimental tests will confirm the FMS framework, and will show that elementary particles like the Higgs are actually composite, extended objects even within the framework of the standard model of particle physics. Or, this will show that the current formulation of the standard model of particles as a quantum gauge field theory is insufficient, either on formal grounds or because the model is incomplete. Either way, the decision will change drastically our view of the world.

While even this aspect has been drastically underestimated, the far-reaching consequences of the FMS framework are even more so. The insistence on forcing a manifestly and non-perturbative gauge-invariant approach even at arbitrarily weak coupling and a convenient hiding of the gauge symmetry by the BEH effect shows that it was possibly to take field theory seriously without loosing the technical ability to be predictive. In fact, in view of the Gribov-Singer ambiguity and the theorems of Haag and Elitzur, it provides a much better understanding of why (and when) perturbation theory can be a quantitatively viable approach.

At the same time, this reasoning is a role model. The FMS framework showed how further quantum gauge theories beyond the standard model should be approached: From the question of physical observables, and maintaining formal consistency. Approximations need to maintain consistency to a much better degree as standard perturbation theory does, which even in non-gauge theories runs afoul of Haag's theorem. Applications beyond the standard model showed explicitly that results based on the FMS framework are in much better agreement with full, non-perturbative results, even at very weak coupling, than those which break formal consistency like perturbation theory.

Moreover, the FMS framework shows that what is usually called confinement is not a distinct phenomena, but could really be viewed as an aspect of manifest gauge invariance \cite{Lavelle:1995ty}. This unifies the way how physical observable particles in the standard model should be treated, and removes the necessity to separate between the strong interaction and the electroweak one in conceptual terms \cite{Maas:2017wzi}. This generalizes then to arbitrary other theories, up to and including quantum gravity ones. Especially, it implies that any gauge theory needs to be considered ontologically to be a theory of extended objects, rather than point-like elementary particles. Confirming this in the standard-model case experimentally would indeed change disruptively the way how we think about the laws of nature. Thus, Giovanni Morchio's legacy could very well become a crucial stepping stone in particle physics and the search for the most fundamental laws of nature and what reality is.

\begin{acknowledgement}
I am grateful to P.~Berghofer and S.~Pl\"atzer for a critical reading of the manuscript. Part of this work has been supported by the the Austrian Science Fund FWF, grant
P32760 and doctoral school grant W1203-N16. Part of the results have been obtained using the HPC clusters at the University of Graz and the Vienna Scientific Cluster (VSC).\\

This is a preprint of the following chapter: Axel Maas, ``The Fr\"ohlich-Morchio-Strocchi mechanism: A underestimated legacy'', published in ``Trails in Modern Theoretical and Mathematical Physics '', edited by Andrea Cintio and Alessandro Michelangeli, 2023, published by Springer, Cham, reproduced with permission of Springer Nature Switzerland AG 2023. The final authenticated version is available online at: http://dx.doi.org/10.1007/978-3-031-44988-8.
\end{acknowledgement}

\bibliographystyle{spphys.bst}
\bibliography{bib.bib}

\end{document}